\documentclass[final,5p,times,twocolumn]{elsarticle} 

\usepackage{graphicx}

\usepackage{amssymb}

\usepackage{color}

\journal{Nuclear Instruments and Methods A}

\begin{document}

\begin{frontmatter}



\title{The Assembly of the Belle II TOP Counter}

\author{Boqun Wang, On behalf of the Belle II PID Group}
\ead{boqunwg@ucmail.uc.edu}
\address{Department of Physics, University of Cincinnati, Cincinnati, OH, USA\\
{\rm University of Cincinnati preprint UCHEP-14-01} }


\begin{abstract}

A new type of ring-imaging Cherenkov counter, called TOP
counter, has been developed for particle identification at the
Belle II experiment to run at the SuperKEKB accelerator in KEK, Japan.
The detector consists of 16 identical modules arranged azimuthally
around the beam line. The assembly procedure for a TOP module is
described. This procedure includes acceptance testing of the quartz
mirror, prism, and quartz bar radiators. The acceptance tests
include a chip search and measurements of bulk transmittance
and total internal reflectance. The process for aligning and
gluing the optical components together is described.

\end{abstract}


\end{frontmatter}

\section{Introduction}

\begin{figure}[htb]
\centerline{
\includegraphics[width=\columnwidth]{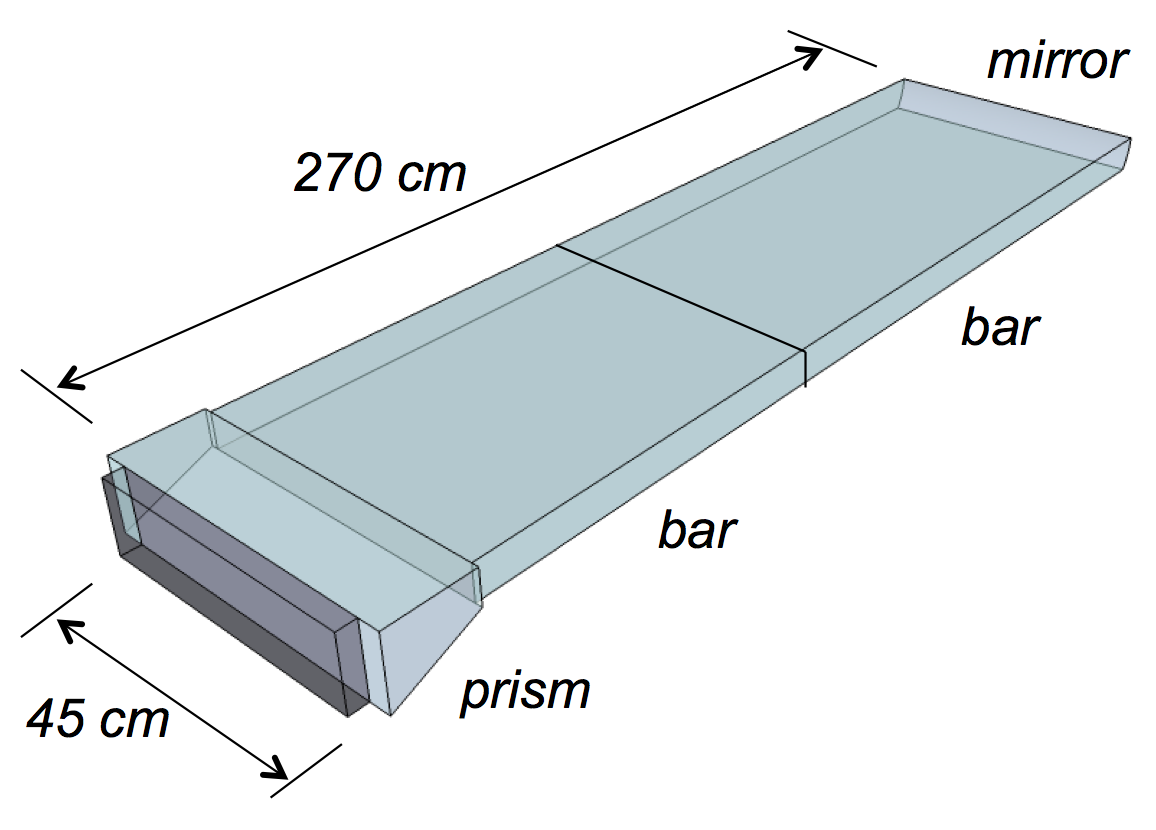}}
\caption{Overview of the TOP counter. The length for one module is 270 cm and the width is 45 cm. It consists of two 125 cm long quartz bars, one 10 cm long mirror with focal length as 5 m, and one 10 cm long prism with a $\sim18^{\circ}$ tilted angle}
\label{overview}
\end{figure}

The time of propagation detector named TOP counter~\cite{top-refs1,top-refs2,top-refs3,top-refs4,top-refs5,top-refs6} is the particle identification detector developed for the Belle II/SuperKEKB experiment~\cite{belle-ii}. As shown in Figure~\ref{overview}, one TOP module consists 2.7 m long quartz bar radiator with a spherical mirror on one end and a prism wedge on the other end. The time of propagation of the Cherenkov photons traveling inside the radiator is detected by an array of micro-channel-plate photomultiplier tubes (MCP-PMTs)~\cite{mcp-pmt1,mcp-pmt2} attached to the prism. The Cherenkov ring image is reconstructed as the pattern of the detected time and position of the Cherenkov photons. The quartz bar radiator must transmit the Cherenkov photons over long optical length with a large number of internal reflections. Therefore, high bulk transmittance and total internal reflectance are required. The quartz bars should be optically uniform, and then the level of damage on the surfaces and edges must be low. The goal being that the photon detectors detect as many photons as possible to reach the design $K/\pi$ separation power.

A total of 16 identical TOP modules will be installed on the Belle II detector. Therefore 32 quartz bars, 16 mirrors and 16 prisms should be manufactured, tested and glued together. In this paper, we will describe the fabrication process of the TOP counter; the acceptance test results for the quartz bars, mirrors and prisms after their delivery, which includes the bulk transmittance and internal reflectance for bars, the tilted angle for prism and the focal length and reflectivity for mirror; and the alignment and gluing procedure to assemble all optical components together for the following bar box sealing.

\section{Fabrication process}

The optical components, which includes quartz bars, prisms and mirrors, are manufactured by several vendors. The surface flatness, surface roughness, parallelism, perpendicularity and chamfer specs will be measured and guaranteed by vendors. The metrology report containing the properties above is provided with the optical components.

The quartz bars will be delivered directly to KEK, Japan. After delivery, the surface and edge quality will be roughly checked by eyes then carefully checked by an automatic chip search system. If the size and number of the chips and scratches satisfy the specification, the bar will be accepted. Then the bulk transmittance and internal reflectance of the bar will be measured by the automatic measurement system. The measurement results will be used for MC simulation.

The mirrors and prisms will be delivered to University of Cincinnati, USA. Their surface qualities will be checked first like the quartz bars. The transmission and reflectivity will be measured, too. For mirrors, additional properties like the position of optical axis, focal point, focal length, spherical aberration and astigmatism will be measured. For prisms, the angle of the tilted surface is a very important property and will be measured carefully. After the above acceptance tests, mirrors and prisms will be packed and delivered to KEK, Japan.

In KEK, after receiving all the optical components, including two bars, one mirror and one prism, they will be mounted on some specially designed gluing stages. The level and angle of these stages are controlled by some micrometers, so these components could be aligned very precisely. With autocollimator and laser displacement sensor, the angle and level difference between two components could be very small.

After the alignment, the gluing process can start. Some special jigs are designed to help the gluing. This includes an air dispensing system, and a mechanical gluing trolley to translate the syringe. The adhesive for gluing is NOA63, which is UV adhesive. 

\begin{figure*}[htb]
\centerline{
\includegraphics[width=0.8\linewidth]{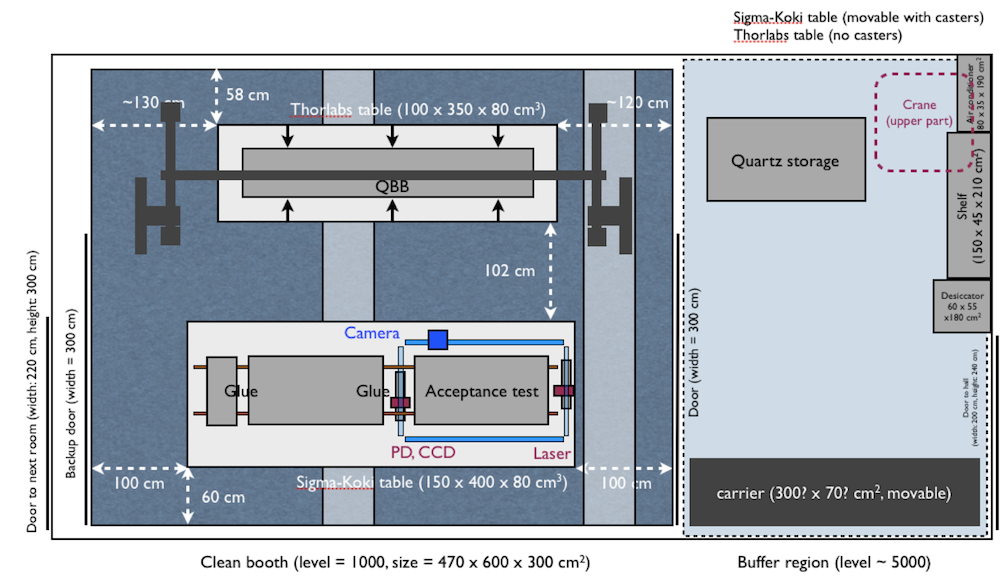}}
\caption{The TOP module assembly clean room in KEK. Outside is a class 5000 buffer room and inside is a class 1000 dark booth. There are two optical tables inside the clean room: one for testing, aligning and gluing, and the other one for QBB assembly. A crane with a specially designed handling jig could move the TOP module from one table to another.}
\label{kek-cleanroom}
\end{figure*}

The assembly room in KEK is shown in Figure~\ref{kek-cleanroom}. This is a clean room with a class 5000 buffer room outside and a class 1000 dark booth inside. Two optical tables stay inside the dark booth. The larger one is used for the testing, aligning and gluing of the optical components. The gluing stages are mounted on two high precision rails which are fixed on the table. After the gluing, the completed TOP module will be moved to another optical table by a specially designed handling jig and crane. Then the module will be sealed in the quartz bar box (QBB). Finally, the TOP module will be moved outside the clean room for mounting electronics and cosmic ray testing.

\section{Acceptance tests}

By the end of February, 2014, two bars were delivered to KEK (one made by Zygo and the other one made by Aperture-Okamoto). Four prisms (all made by Zygo) and one mirror (made by ITT) were delivered to University of Cincinnati. Some tests are performed with these components and in this section some preliminary results will be shown.

\subsection{Quartz bar}

\begin{figure}[htb]
\centerline{
\includegraphics[width=0.8\columnwidth]{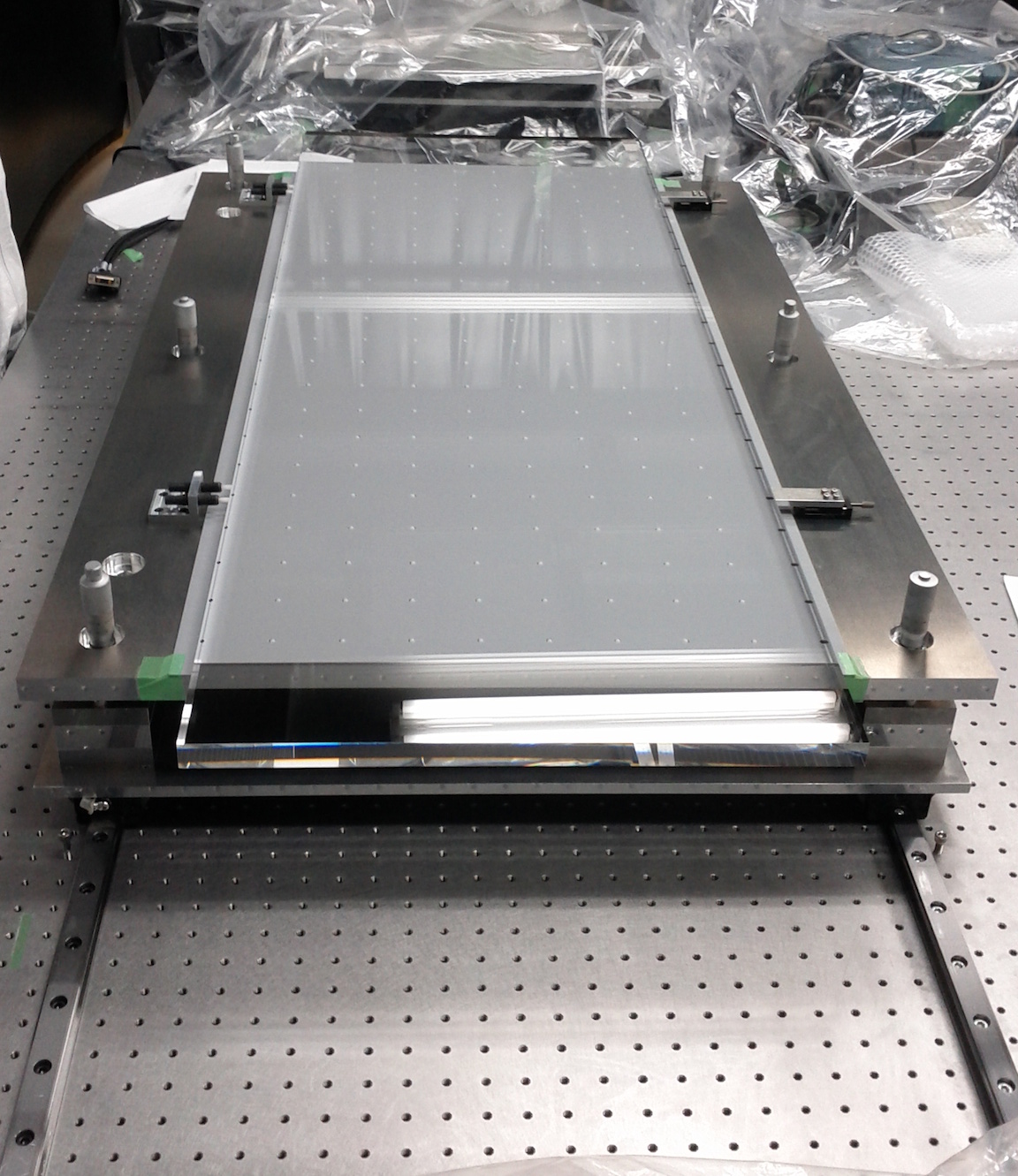}}
\caption{The quartz bar made by Zygo. The bar is mounted on the gluing stage with six micrometers. The gluing stage is mounted on high precision rails which are fixed on the optical table.}
\label{zygo-bar}
\end{figure}

\begin{figure}[htb]
\centerline{
\includegraphics[width=\columnwidth]{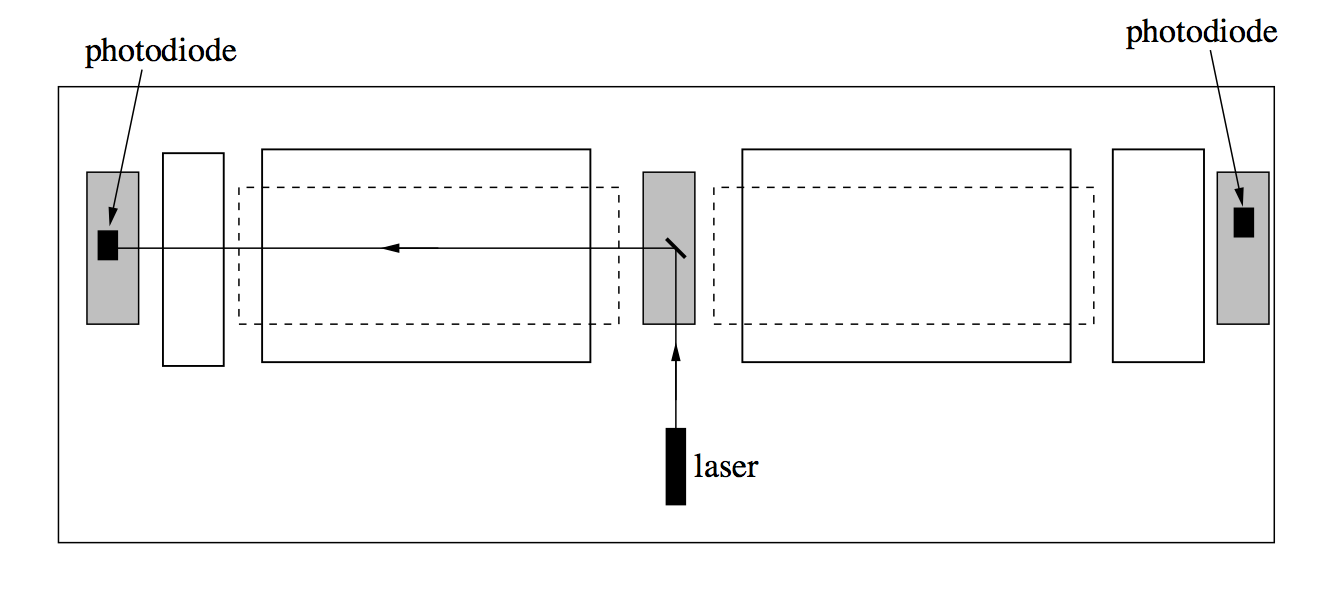}}
\caption{The acceptance test system for quartz bar. Two bars are put on the stages and the x-y stages with laser and photo-diode are installed at the two ends of the bars.}
\label{bar-test}
\end{figure}

The quartz bar made by Zygo is shown in Figure~\ref{zygo-bar}. The acceptance test system for quartz bar is shown in Figure~\ref{bar-test}. The quartz bars are put on the stages and the measurement system is installed around them. The laser and photo-diode are on the two ends of the quartz bars, and they are used to measure the bulk transmittance and internal reflectance of the quartz bars.The position of the laser and photo-diode is controlled by x-y stages, which are controlled by a notebook computer. The measurement system is automatic, which means after setting up the system, the measurement will be finished automatically without human intervention. After optimization, the scan speed is quite fast. We can finish the scanning of $\sim$400 points within 5 h.

\begin{figure}[htb]
\centerline{
\includegraphics[width=0.5\columnwidth]{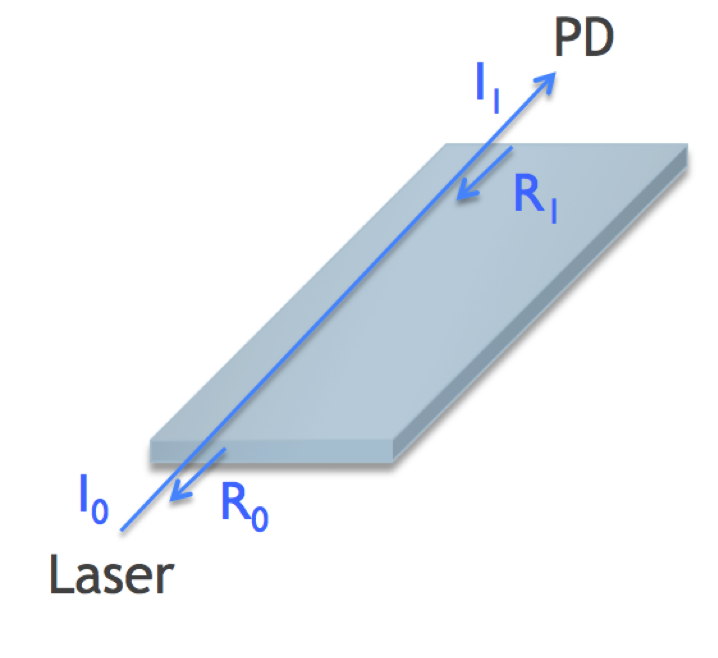}
\includegraphics[width=0.5\columnwidth]{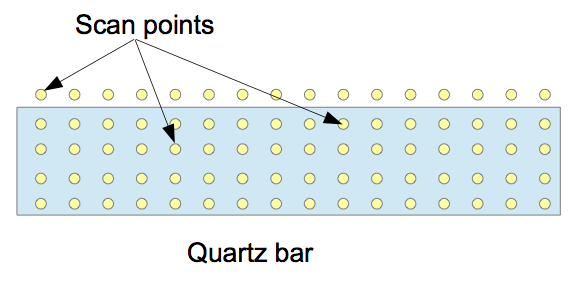}}

\centerline{
\includegraphics[width=1.0\columnwidth]{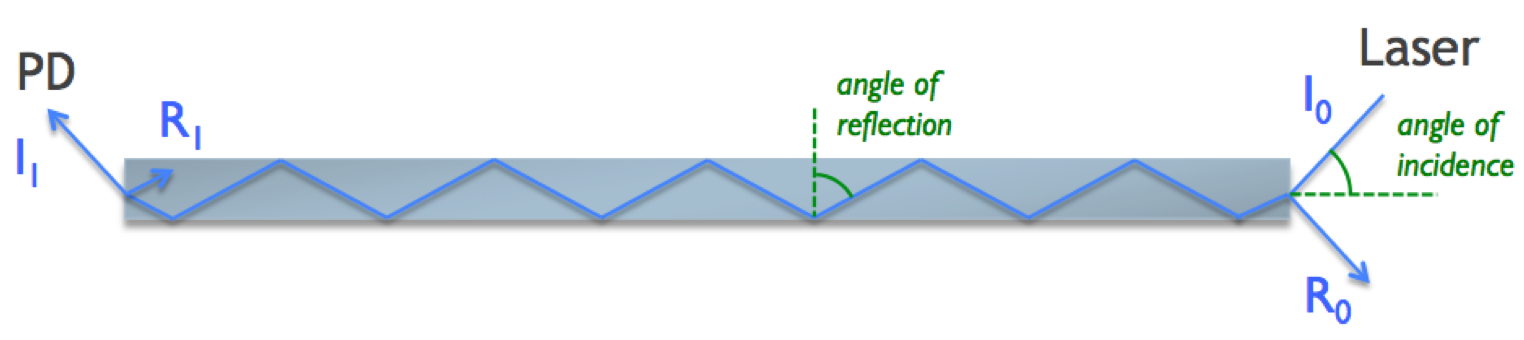}}
\caption{The methods of the acceptance tests. On the top is the method for bulk transmittance and the scan point distribution. On the bottom is the method for internal reflectance measurement.}
\label{bar-acc-method}
\end{figure}

The methods for bulk transmittance and internal reflectance measurement are shown in Figure~\ref{bar-acc-method}. The scanning points are also shown in Figure~\ref{bar-acc-method}. 

The bulk transmittance $\tau$ is obtained from the equation
\begin{equation}
	I_0 (1-R_0) \tau (1-R_1) = I_1
\end{equation}
where $I_0$ is the intensity of the incident laser, $I_1$ is the intensity after the transmission through the quartz bar, and $R_0$ and $R_1$ are the reflectance at the air-to-bar and bar-to-air surfaces respectively, which could be calculated by Fresnel equation. The wavelength of the laser we used in this and the following measurements are 405 $nm$.

The internal reflectance $\alpha$ is obtained from the equation
\begin{equation}
\label{eq:ir}
	I_0 (1-R_0) \alpha^N e^{-L/\Lambda\sqrt{1+\left(bN/L\right)^2}} (1-R_1) = I_1
\end{equation}
where $I_0$, $I_1$, $R_0$ and $R_1$ have the same definitions as bulk transmittance. $N$ is the number of bounces, $\Lambda$ is the attenuation length defined as the path length in the bulk region corresponding to the photon retainment of $1/e$ which could be calculated from the result of the bulk transmittance measurement, and $L$ and $b$ are the length and the thickness of the bar, respectively. The exponential factor indicates the photon loss due to the bulk transmission.

\begin{figure*}[htb]
\centerline{
\includegraphics[width=0.47\linewidth]{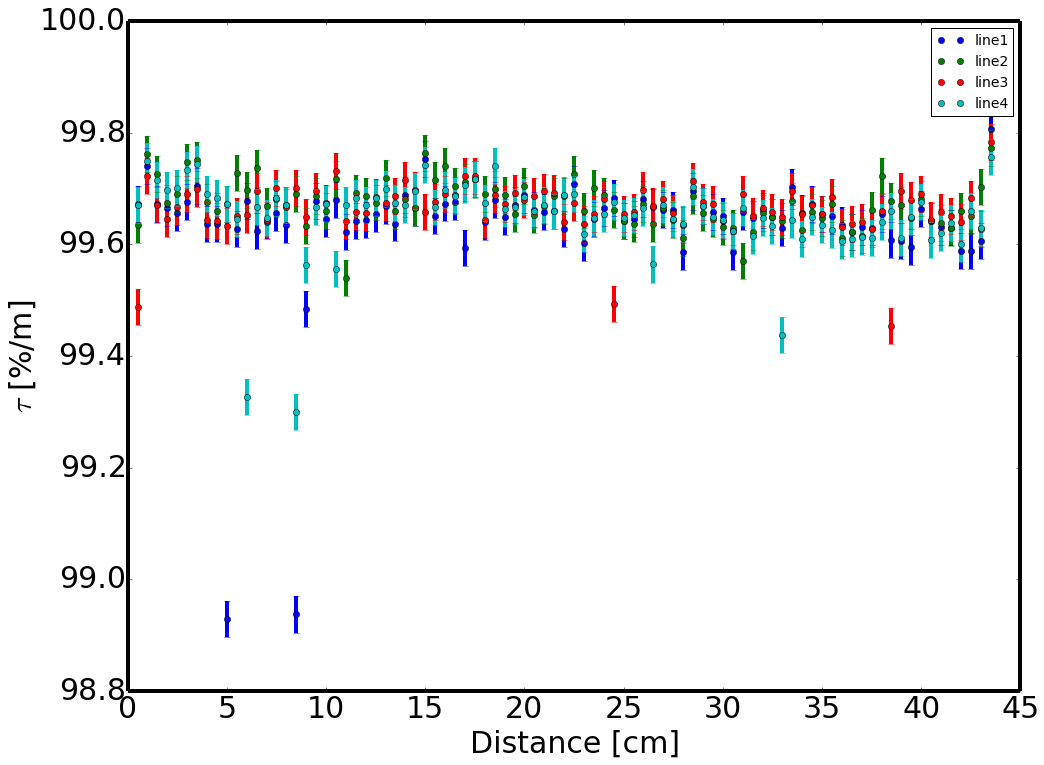}
\includegraphics[width=0.47\linewidth]{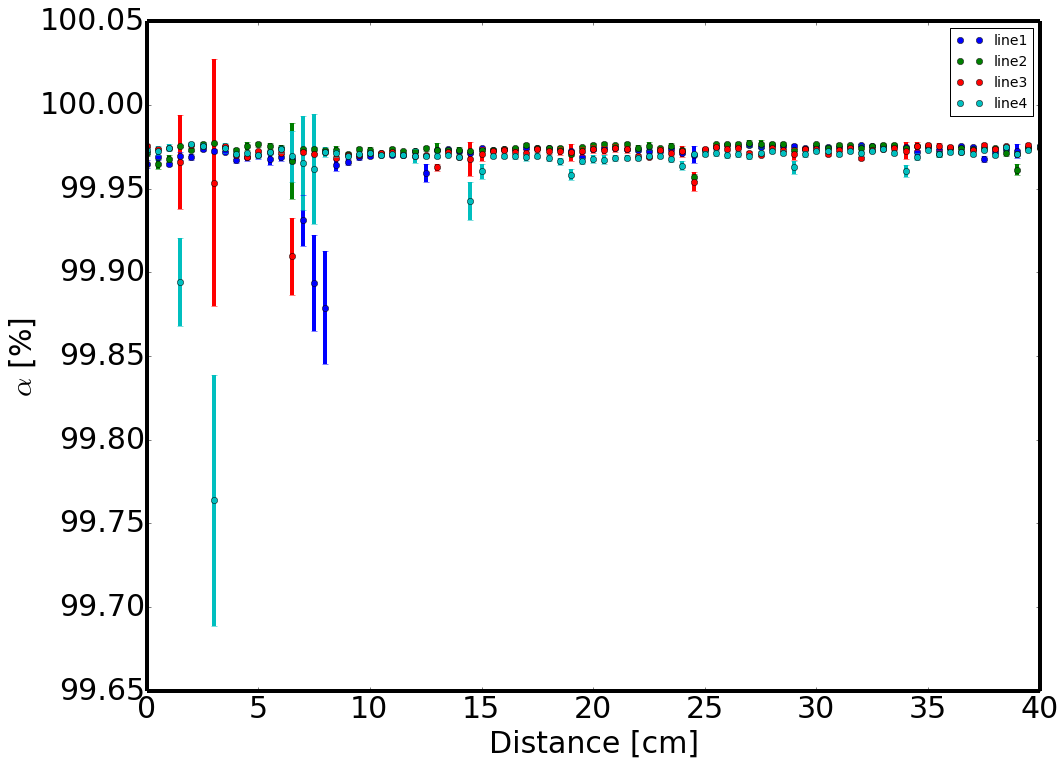}}
\caption{The results of the bulk transmittance (left) and internal reflectance with $N=21$ in Equation \ref{eq:ir} (right). Four different colors correspond to four scanning lines inside the quartz bar. The large uncertainty near 10 cm is caused by the scratch on the surface.}
\label{bar-acc-result}
\end{figure*}

The results of the quartz bar acceptance tests are shown in Figure~\ref{bar-acc-result}. For the zygo bar, the bulk transmittance measurement result is ($99.66 \pm 0.07$)\%/m, and the internal reflectance measurement result is ($99.971 \pm 0.013$)\%. Some scratches on the quartz surface may affect the result. The results for Aperture-Okamoto bar are ($99.54 \pm 0.08$)\%/m for bulk transmittance and ($99.982 \pm 0.013$)\% for internal reflectance. The errors contain both statistical and systematic errors, which include the contribution from bulk transmittance measurement, laser stability, angle measurement and laser polarization.

\subsection{Mirror and prism}

\begin{figure}[htb]
\centerline{
\includegraphics[width=0.8\columnwidth]{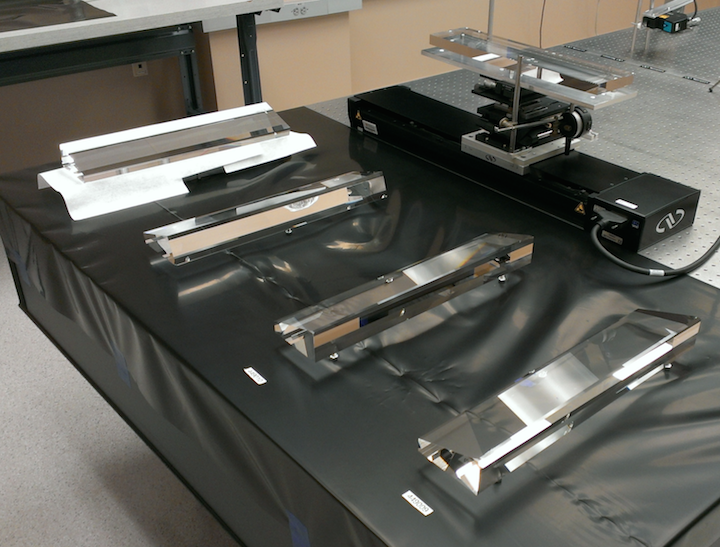}}
\caption{The mirror and prism. The mirror is mounted on stage and the four prisms are on the optical table.}
\label{mirror-prism}
\end{figure}

The first mirror made by ITT and four prisms made by Zygo have been delivered to University of Cincinnati, as shown in Figure~\ref{mirror-prism}. The surface quality is good, and only a few small chips are found.

\begin{figure*}[htb]
\centerline{
\includegraphics[width=\linewidth]{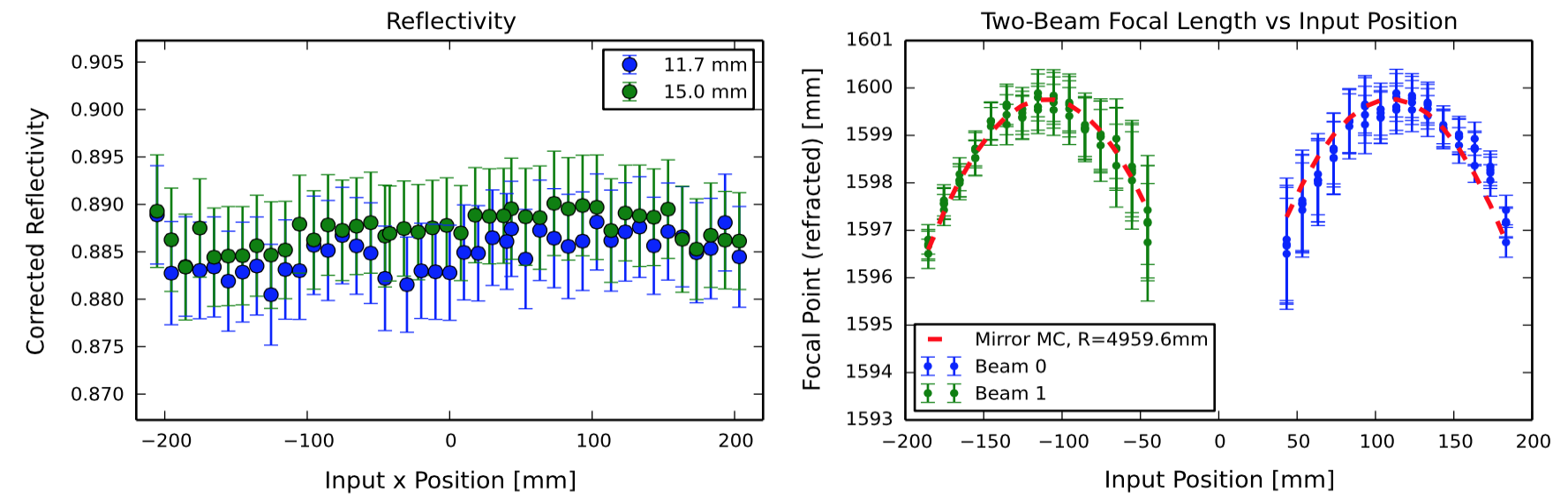}}
\caption{The testing result of mirror reflectivity (left) and focal length (right).}
\label{mirror-result}
\end{figure*}

The mirror has been tested for reflectivity and focal length. The result of the mirror tests is shown in Figure~\ref{mirror-result}. The reflectivity is $(88.51 \pm 0.20)$\% for the data taken at $11.7$ mm from surface S1, and $(88.73 \pm 0.18)$\% for the data taken at $15.0$ mm from surface S1. The focal length is $(4965.35 \pm 11.84)$ mm for the $x > 0$ half of the mirror and $(4967.55 \pm 12.55)$ mm for the $x < 0$ half of the mirror. The laser wavelength is $532$ nm.


The testing of the prism is still on going. The preliminary result of the tilted angle is $(18.085 \pm 0.003)^{\circ}$. This result is consistent with the specification which is $(18.07 \pm 0.04)^{\circ}$ and the vendor's measurement which is $18.07^{\circ}$.

\section{Gluing}

After testing, all components will be mounted on specially designed gluing stages with micrometers which can control the level and angle of the stages for alignment. Autocollimator and laser displacement sensor are used to monitor the level and angle difference between them. 

\begin{figure}[htb]
\centerline{
\includegraphics[width=\columnwidth]{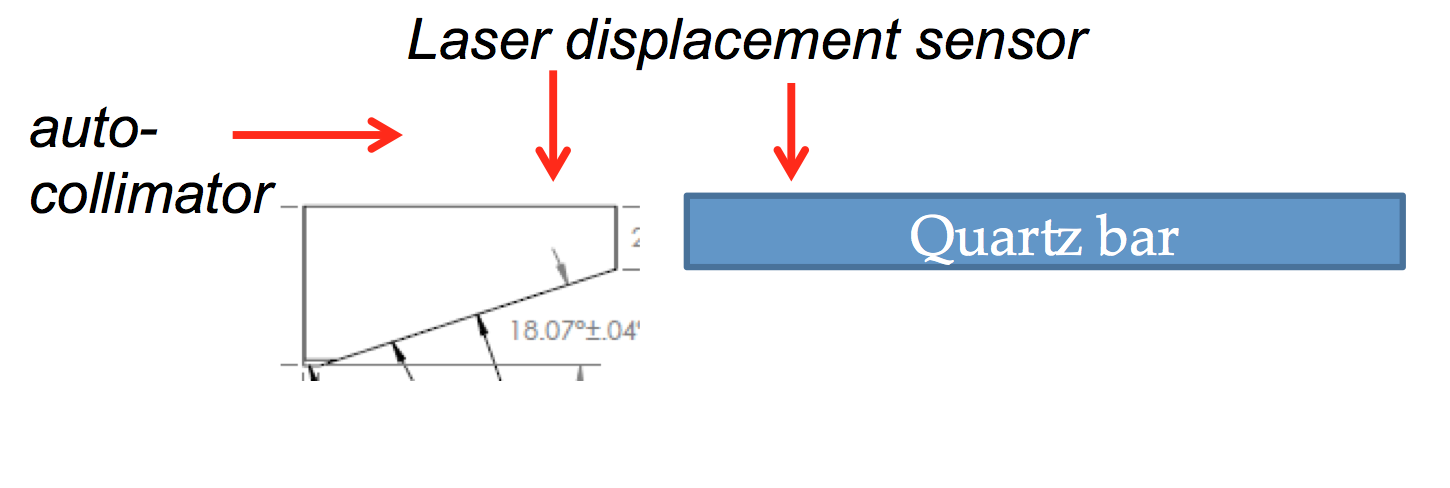}}
\caption{The alignment for gluing. This shows the condition for prism-bar gluing. For bar-bar and bar-mirror gluing the setup is similar.}
\label{gluing-align}
\end{figure}

After the alignment, the optical components will be glued together with adhesives. The adhesive for the gluing is NOA63, which is UV adhesive. For one TOP module, there are three kinds of gluing: bar to bar, bar to mirror and bar to prism. For all of them, the gluing setup is similar, which is shown in Figure~\ref{gluing-align}.


Tests for the gluing have been taking place. A specially designed mechanical trolley was used and proved to be useful for easing up the gluing. A few tests succeeded with barely no visible bubbles and others ended with some large and small bubbles. We tested the gluing procedure with different gaps between the optical components. If the gap is too small, the adhesive will flow very slowly and it is very sensitive to the surface roughness and cleanliness. This may generate lots of bubbles. More tests are needed before the assembly of the first TOP counter.


\section{Summary}

The TOP counter is important for Belle II detector's particle identification capability. The mass production of the optical components has started and the basic assembly procedure has been fixed. After delivery, the four components (two quartz bars, one prism and one mirror) will be tested separately in the University of Cincinnati and KEK. Then they will be aligned and glued together in KEK and ready for QBB assembly.

For now, there are two quartz bars, four prisms and one mirror which are manufactured and delivered by vendors. Their properties are measured and the preliminary results show that they satisfy our specification. The assembly of the first TOP counter will start soon. Before that, further tests of the gluing procedure are needed.


%

\end{document}